\documentclass[prx,reprint,twocolumn,longbibliography]{revtex4-1}
\usepackage{amsmath,amssymb,bm,mathrsfs,graphicx, braket, times,amsthm,enumerate, scrextend}
\usepackage[colorlinks=true,citecolor=blue,linkcolor=blue,urlcolor=blue]{hyperref}
\usepackage{longtable}
\usepackage{multirow}
\usepackage{array}
\usepackage{bigstrut}
\usepackage[all,cmtip]{xy}
\usepackage[normalem]{ulem}
\usepackage[usenames,dvipsnames]{color}
\usepackage{makecell}
\usepackage{xcolor}
\usepackage{float}
\usepackage{graphicx}

\begin{document}

\title{Novel three-dimensional Fermi surface and
electron-correlation-induced charge density wave in FeGe}
\author{Lin Wu}
\author{Yating Hu}
\author{Di Wang}
\email{diwang0214@nju.edu.cn}
\author{Xiangang Wan}
\email{xgwan@nju.edu.cn}
\affiliation{National Laboratory of Solid State Microstructures and School of Physics,
Nanjing University, Nanjing 210093, China}
\affiliation{Collaborative Innovation Center of Advanced Microstructures, Nanjing
University, Nanjing 210093, China}

\begin{abstract}
As the first magnetic kagome material to exhibit the charge density wave
(CDW) order, FeGe has attracted much attention in recent studies. Similar to
AV$_{3}$Sb$_{5}$ (A = K, Cs, Rb), FeGe exhibits the CDW pattern with an
in-plane 2$\times $2 structure and the existence of van Hove singularities
(vHSs) near the Fermi level. However, sharply different from AV$_{3}$Sb$_{5}$
which has phonon instability at $M$ point, all the theoretically calculated
phonon frequencies in FeGe remain positive. Here, we perform a comprehensive
study of the band structures, Fermi surfaces, nesting function and the
mechanism of CDW transition through first-principles calculations.
Surprisingly, we find that the maximum of nesting function is at $K$ point
instead of $M$ point. Two Fermi pockets with Fe-$d_{xz}$ and Fe-$%
d_{x^{2}-y^{2}}$/$d_{xy}$ orbital characters have large contribution to the
Fermi nesting, which evolve significantly with $k_{z}$, indicating the
highly three-dimensional (3D) feature of FeGe in contrast to AV$_{3}$Sb$_{5}$%
. Meanwhile, the vHSs are close to the Fermi surface only in a small $k_{z}$
range, and does not play a leading role in nesting function. Considering the
effect of local Coulomb interaction, we reveal that the Fermi level
eigenstates nested by vector $K$ are mainly distributed from unequal
sublattice occupancy, thus the instability at $K$ point is significantly
suppressed. Meanwhile, the wave functions nested by vector $M$ have many
ingredients located at the same Fe site, thus the instability at $M$ point
is enhanced. This indicates that the electron correlation, rather than
electron-phonon interaction, plays a key role in the CDW transition at $M$
point.
\end{abstract}

\date{\today}
\maketitle
\date{\today }

Kagome lattice materials \cite%
{kagome_lattice,yin2022topological,Chen_2022,review_02061} exhibit unique
electronic structure signatures owing to their unconventional geometric
characteristics, embracing the flat bands induced by destructive
interference of nearest-neighbour hopping, a pair of van Hove singularities
(vHs) at $M$ point, and Dirac cone dispersion at $K$ point \cite%
{kagome-04,kagome-05,PhysRevLett.100.136404,kagome-03,kagome-06,phase2,kagome3band,PhysRevB.80.113102}%
. A wide array of exotic physical phenomena in the kagome lattice arises
from different degrees of electron filling. When large density of states
(DOS) from the kagome flat bands are located near the Fermi level, strong
electron correlations can induce magnetic order \cite%
{kagome-04,kagome-05,PhysRevLett.100.136404}. Meanwhile, when vHSs are
located near the Fermi level, interaction between the saddle points and
lattice instability could induce charge density wave (CDW) order with a 2$%
\times $2 structure in xy-plane \cite%
{kagome-03,kagome-06,phase2,kagome3band,PhysRevB.80.113102}. Therefore
kagome lattice materials serve an essential platform for studying non-trival
topological physics \cite{kagome_topo,Mn3Sn,Co3Sn2S2_1,Co3Sn2S2_2,Co3Sn2S2_3}%
, CDW order \cite%
{135_new,PhysRevMaterials.6.015001,miao2021geometry,subedi2022hexagonal,135-05,135-09,135-11,135_nature2,135_PRL2,135_nature3,135_NC1,135_NC2,135_PRM1,135_nest,135_NP2,135_EPC1,135_EPC2,zheng2022emergent,135_hjp1,135_hjp2,135_hjp3,phase3}%
, superconductivity \cite{135_nature0,135_PRL1,135_PRM2,135-01,jeong2022crucial}, fractional
quantum Hall effect \cite{kagome_QHE,PhysRevB.83.165118,PhysRevB.81.235115}
and quantum\ anomalous Hall effect (QAHE) \cite%
{kagome_AHE1,kagome_AHE2,Mn3Ge_1,Mn3Ge_2,135_QAHE1,135_QAHE2}.

As a well-known family of non-magnetic layered kagome compounds, AV$_{3}$Sb$%
_{5}$ (A = K, Cs, Rb) \cite{135_new} were reported to host CDW \cite%
{PhysRevMaterials.6.015001,miao2021geometry,subedi2022hexagonal,135-05,135-09,135-11,135_nature2,135_PRL2,135_nature3,135_NC1,135_NC2,135_PRM1,135_nest,135_NP2,135_EPC1,135_EPC2,zheng2022emergent}%
, superconductivity \cite{135_nature0,135_PRL1,135_PRM2,135-01,jeong2022crucial}, giant QAHE
\cite{135_QAHE1,135_QAHE2} and chiral flux phase \cite%
{135_hjp1,135_hjp2,135_hjp3}. In this system, vHSs are located near the
Fermi level and the phonon spectrum exhibit two imaginary phonon frequencies
locating at the Brillouin zone (BZ) boundary ($M$ ($\frac{1}{2},0,0$) and $L$
($\frac{1}{2},0,\frac{1}{2}$) points) \cite{135_PRM1,135_EPC1}, which induce
an in-plane 2$\times $2 CDW state identified in experiment \cite%
{135-05,135-09,135-11,135_nature2,135_PRL2,135_nature3}. On the other hand,
there are many kagome magnetic materials such as FeSn \cite%
{FeSn1,FeSn2,FeSn3,FeSn4,FeSn5}, Fe$_{3}$Sn$_{2}$ \cite{Fe3Sn2_1,Fe3Sn2_2},
Mn$_{3}$Sn \cite{Mn3Sn}, Mn$_{3}$Ge \cite{Mn3Ge_1,Mn3Ge_2} and Co$_{3}$Sn$%
_{2}$S$_{2}$ \cite{Co3Sn2S2_1,Co3Sn2S2_2,Co3Sn2S2_3}. However, it may be due
to the large energy separation between flat bands and vHSs that CDW order
and magnetic order have not been generally observed simultaneously in one
material \cite{FeGe_exp1,Zhj}.

Very recently, the discovery of CDW order coexists with magnetic order in
the Kagome material FeGe was reported from a joint experimental study of
angle-resolved photoemission spectroscopy (ARPES) \cite{FeGe_exp1,FeGe_exp2}%
, neutron and x-ray scattering \cite{FeGe_exp2,FeGe_exp3,FeGe_EPC}, scanning
tunneling microscopy (STM) \cite{FeGe_exp2,FeGe_exp3,FeGe_cdw2,2302.04490}
and photoemssion \cite{FeGe_exp3}, which offers a fascinating platform to
investigate interplay between the CDW and magnetism. FeGe exhibits a
sequence of phase transitions: (1) an A-type antiferromagnetism (AFM) phase
appears below the high magnetic transition temperature up to $T_{c}\sim $
410K \cite{FeGe_1963}, (2) a CDW phase \cite%
{FeGe_exp2,FeGe_exp3,FeGe_EPC,FeGe_cdw2} takes place at $T_{CDW}\sim $ 100K
and the system becomes stable in 2$\times $2$\times $2 superstructure as
previously reported in metal AV$_{3}$Sb$_{5}$ \cite%
{135-05,135-09,135-11,135_nature2,135_PRL2,135_nature3}, and (3) the
magnetic moment weakly cants along the c direction, giving a c-axis double
cone AFM structure below a lower transition temperature $T_{canting}$ around
60K \cite{FeGe_1972,FeGe-1975,FeGe_1977,FeGe-1978, FeGe_1984,FeGe_1988}.
Subsequent reports reveal that the vHSs in FeGe, which are originally far
from Fermi surface in a nonmagnetic system, are brought to the vicinity of
Fermi level due to spin exchange splitting of about 1.8eV caused by magnetic
order \cite{FeGe_exp1,FeGe_exp2,Zhj}. Though FeGe exhibits the existence of
vHSs at the $M$ point near the Fermi level similar to AV$_{3}$Sb$_{5}$, it
is note-worthy that the imaginary phonon around the $M$ point present in AV$%
_{3}$Sb$_{5}$ \cite{135_PRM1,135_EPC1} is not observed in the theoretically
calculated phonon spectrum of FeGe \cite{FeGe_exp1,FeGe_cdw2,FeGe_EPC}.
Therefore, a comprehensive study of the band structures, Fermi surfaces,
nesting function, and the mechanism of CDW transition in kagome FeGe is an
emergency issue, which we will address in this work based on first
principles study.

In the present work, we perform a detailed analysis of the electronic
structure and an investigation on the nesting function of A-type AFM kagome
materials FeGe by employing first-principles calculations. We analyse the
variation of the energy band structure and the Fermi surface for different $%
k_{x}$-$k_{y}$ planes as $k_{z}$ changes from 0 to 0.5, and find that the
electronic structures of FeGe have strong 3D feature. The positions of vHSs
evidently shift as $k_{z}$ varies, and only in a small $k_{z}$ range the
vHSs are in proximity to Fermi level ($\pm $0.1 eV), distinct from the
quasi-two-dimensional structural characteristics of AV$_{3}$Sb$_{5}$ \cite%
{135_EPC1,135_nest}. Our numerical results show that the maximum of nesting
function is at the $K$ point instead of the $M$ point. Whereafter, we find
that two pockets have large contribution to the nesting function, which
respectively contribute form the $d_{xz}$ orbital and a combination of $%
d_{x^{2}-y^{2}}$/$d_{xy}$ orbitals. To understand the conflict between the
nesting function at the $K$ point and CDW transition at $M$ point, we
consider the effect of local Coulomb interaction \cite{green_function}. We
find that the Fermi surface nesting at the $K$ point is ineffective due to
different sublattice characters of the band structures, similar with the
sublattice mechanism in superconductors \cite{kagome3band,sub-prl}. On the
other hand, the CDW instability at the $M$ point could be enhanced by the
local Coulomb interaction, since the wave functions nested by vector $M$ are
mainly distributed from the same Fe site. It implies that the strong 3D
Fermi surfaces\ and local electron correlation play indispensable parts in
the CDW transition in FeGe.

We carried out first-principles density functional theory (DFT) calculations
by employing the full-potential all-electron code Wien2k \cite{WIEN_2020}.
The local spin density approximation (LSDA) \cite{vosko1980accurate} was
used as the exchange-correlation functional in our calculations for the
A-type AFM state. We chose a fine k-mesh of 200$\times $200$\times $100 in
the irreducible BZ to ensure that the nesting function calculated from the
eigenvalues is robust to the number of $K$ points. Since our conclusions are
quite different from the typical kagome materials such as AV$_{3}$Sb$_{5}$
\cite{135_PRM1,135_EPC1,135_nest}, we have also used the Vienna ab initio
Simulation Package (VASP) \cite{VASP_1993,VASP_1996} with the projector
augmented wave (PAW) \cite{paw_1994,paw_1999} method to confirm our
electronic structure calculations. The results of the two methods are well
consistent, and in this paper we present the calculated results from Wien2k.

We start by performing LSDA calculation for FeGe based on the experimental
crystal structure (see Supplemental Materials (SM)) and A-type AFM ground
state \cite{FeGe_1988}. The band structure and the density of states (DOS)
(see SM) are basically similar in many aspects to the isostructure FeSn \cite%
{FeSn1,FeSn2,FeSn3,FeSn4,FeSn5}. Our calculations reproduce the results of
previous studies \cite{FeGe_exp1} quite well. The magnetic moment of the Fe
ions is estimated to be 1.55 $\mu _{B}$, which is close to the previous
experimental value of about 1.72 $\mu _{B}$\cite{FeGe-1975,FeGe-1978}.
Besides, similar to AV$_{3}$Sb$_{5}$ \cite{135_EPC1,135_nest}, we find a
pair of vHSs close to Fermi surface on both the $k_{z}$=0 and 0.5 high
symmetric plane as shown in Fig. \ref{dos} of SM.

In order to analyse the orbital components of the energy bands, it is
important to take the proper local coordinate axis. Since the angle between
the orientations of Fe$_{A}$, Fe$_{B}$, and Fe$_{C}$ (see SM) to the same
nearest neighbour Ge$^{out}$ atom is 120$^{\circ }$ in the global
coordinate, the x/y components of Fe$_{A/B/C}$-$d$ orbitals are not
equivalent. The suitable local coordinate
of Fe$_{A}$, Fe$_{B}$, and Fe$_{C}$ we have chosen are shown in Fig. \ref{structure}(c) of
SM, any two of which could be transformed into each other by the $C_{3}$
symmetry operation. The local x-axis always points to the nearest-neighbor Ge%
$^{in}$ atom and the local z-axis direction is the same as the one in the
global coordinate. In this local coordinate, the five Fe-3$d$ orbitals are
clearly divided into higher $d_{yz}$ and $d_{x^{2}-y^{2}}$ parts and lower $%
d_{z^{2}}$, $d_{xz}$ and $d_{xy}$ parts, consistent with the $e_{g}$-$t_{2g}$
relationship in the ortho-octahedral crystal field. The selection of such
local coordinate is also helpful to the analysis of the Fermi pockets as
shown in the following.

\begin{figure}[tb]
\centering\includegraphics[width=0.5\textwidth]{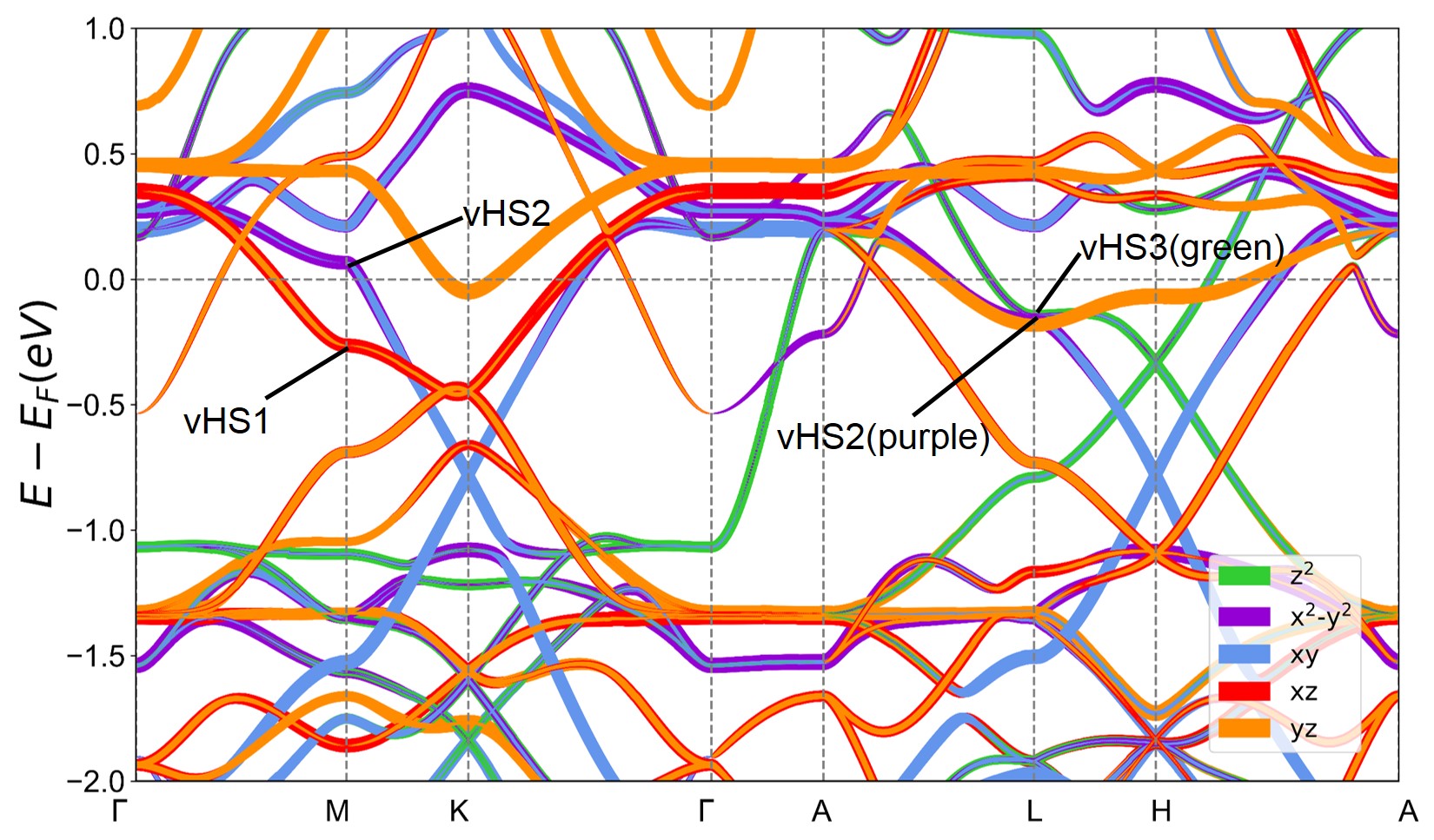}
\caption{The orbital-projected electronic band structure of FeGe near Fermi
level. The Fe-3$d$ orbital projection is performed with the local
coordinate. The orbital characters are labelled by different colors.}
\label{pband00.5}
\end{figure}

We present the orbital components of the energy bands in Fig. \ref{pband00.5}
using the above local coordinate. 
As mentioned above, the kagome lattice with nearest-neighbour hopping will
present the typical three-band structure including flat band, vHS, and Dirac
cone \cite{kagome-04,kagome-05,kagome-03,kagome-06,phase2,kagome3band}.
Along the high symmetry directions $\varGamma -M-K-\varGamma $ lying in $%
k_{z}$=0 plane, there are two different kagome structures near the Fermi
level. One of the Kagome structures consists of Fe-$d_{xz}$ orbitals, which
are located between -2.0 and 0.5 eV, and contain the vHS1 located at -0.26
eV below $E_{F}$, as marked in red in Fig. \ref{pband00.5}. The other kagome
structure containing vHS2 consists of a combination of $d_{x^{2}-y^{2}}$ and
$d_{xy}$ orbitals shown in purple and blue respectively. The vHS2 is located
above the Fermi energy (0.07\ eV). Three Fe-$d_{yz}$ orbitals (labeled in
orange in Fig. \ref{pband00.5}) also form a kagome three-band structure.
However, the interaction between the Fe atoms and the $p_{z}$ orbitals of
the Ge$^{out}$ leads to two bands shifting downward away from $E_{F}$. Along
the $A-L-H-A$ path in the $k_{z}$=0.5 plane there are three kagome
structures. Due to the interaction with the $p$ orbitals of Ge$^{out}$, the
three-band kagome structure composed of Fe-$d_{z^{2}}$ orbitals is located
below and away from the Fermi energy level in the $k_{z}$=0 plane (from
--2.5 to --1.0\ eV), while in the $k_{z}$=0.5 plane the energy bands rise
and cross the Fermi level. Two vHSs at the $L$ point close to the Fermi
energy level are contributed by Fe-$d_{x^{2}-y^{2}}$/$d_{xy}$ and Fe-$%
d_{z^{2}}$ orbitals, marked as vHS2 and vHS3 respectively in Fig. \ref%
{pband00.5} (The analysis below confirms that the vHS2 in $k_{z}$=0 plane
slowly turn into the vHS2 in $k_{z}$=0.5 plane\ with $k_{z}$ shifts from 0
to 0.5). It is worth mentioning that vHSs near $E_{F}$ have different
orbital characters in $k_{z} $=0 and $k_{z}$=0.5 planes, indicating the
important role of $k_{z}$ in electronic structures.

Furthermore, using group theory \cite{Bradley,tang-kp}, we obtain the
irreducible representations (irreps) of the little group for each vHS based
on the wave functions from DFT calculations. At the $M$ point, our
calculations identify the irrep of vHS1 as B$_{3g}$, while vHS2 corresponds
to the irrep A$_{g}$. On the other hand, at $L$ point, the irrep of the vHS2
and VHS3 are both A$_{u}$. The irreps of the vHSs closer to $E_{F}$ are
different in $k_{z}$=0 and $k_{z}$=0.5 planes, also indicating the strong 3D
feature in FeGe, which will be discussed carefully in the following.

\begin{figure}[tb]
\centering\includegraphics[width=0.48\textwidth]{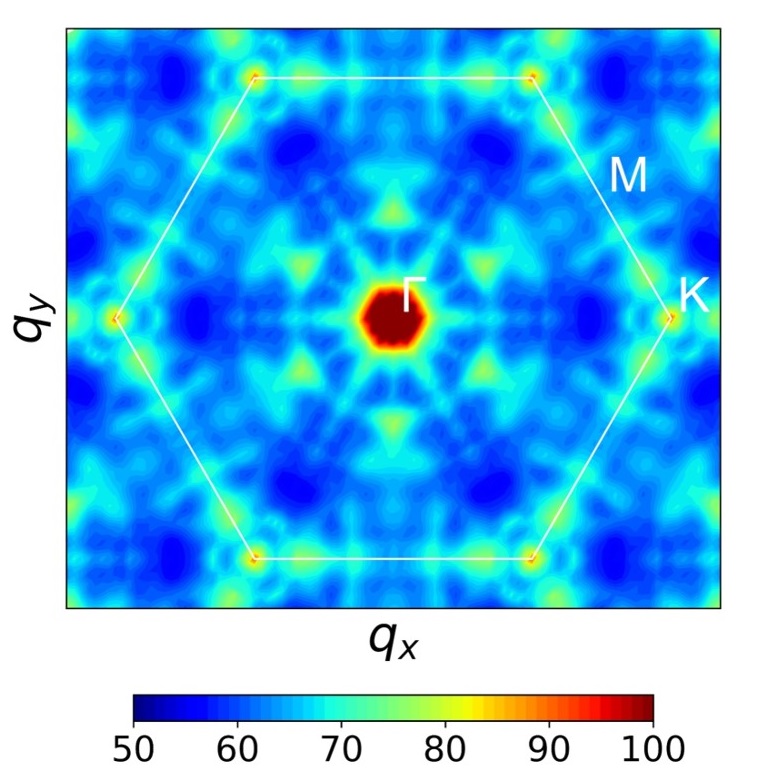}
\caption{The nesting function $\protect\xi (q_{x},q_{y},0)$ of FeGe. We
neglected the peak at $q$=0 to better present the nesting function.}
\label{susall}
\end{figure}

To understand the CDW instability, we calculate the Fermi surface nesting
function $\xi (\mathbf{q})$ \cite{susceptibility}. The 2$\times $2$\times $2
supercell structure of CDW phase for the non-magnetic pristine phase is
suggested by experimental results \cite%
{FeGe_cdw2,FeGe_EPC,FeGe_exp2,FeGe_exp3}. Since the A-type AFM structure has
already enlarged unit cells doubled along the z-direction, we focus the
in-plane $\mathbf{q}$ vector in the following, and obtain the nesting
function $\xi (q_{x},q_{y},0)$, as is illustrated in Fig. \ref{susall}. It
can be seen that the maximum values of nesting function are located at the $%
K $ point instead of the $M$ point, which is different from the results of
the tight-binding model of kagome lattice \cite%
{kagome-03,kagome-06,phase2,kagome3band} and AV$_{3}$Sb$_{5}$ \cite%
{135_EPC1,135_nest}. This motivated a careful analysis of the band
structure, Fermi surface and the nesting function of FeGe.

\begin{figure*}[tbh]
\centering\includegraphics[width=1\textwidth]{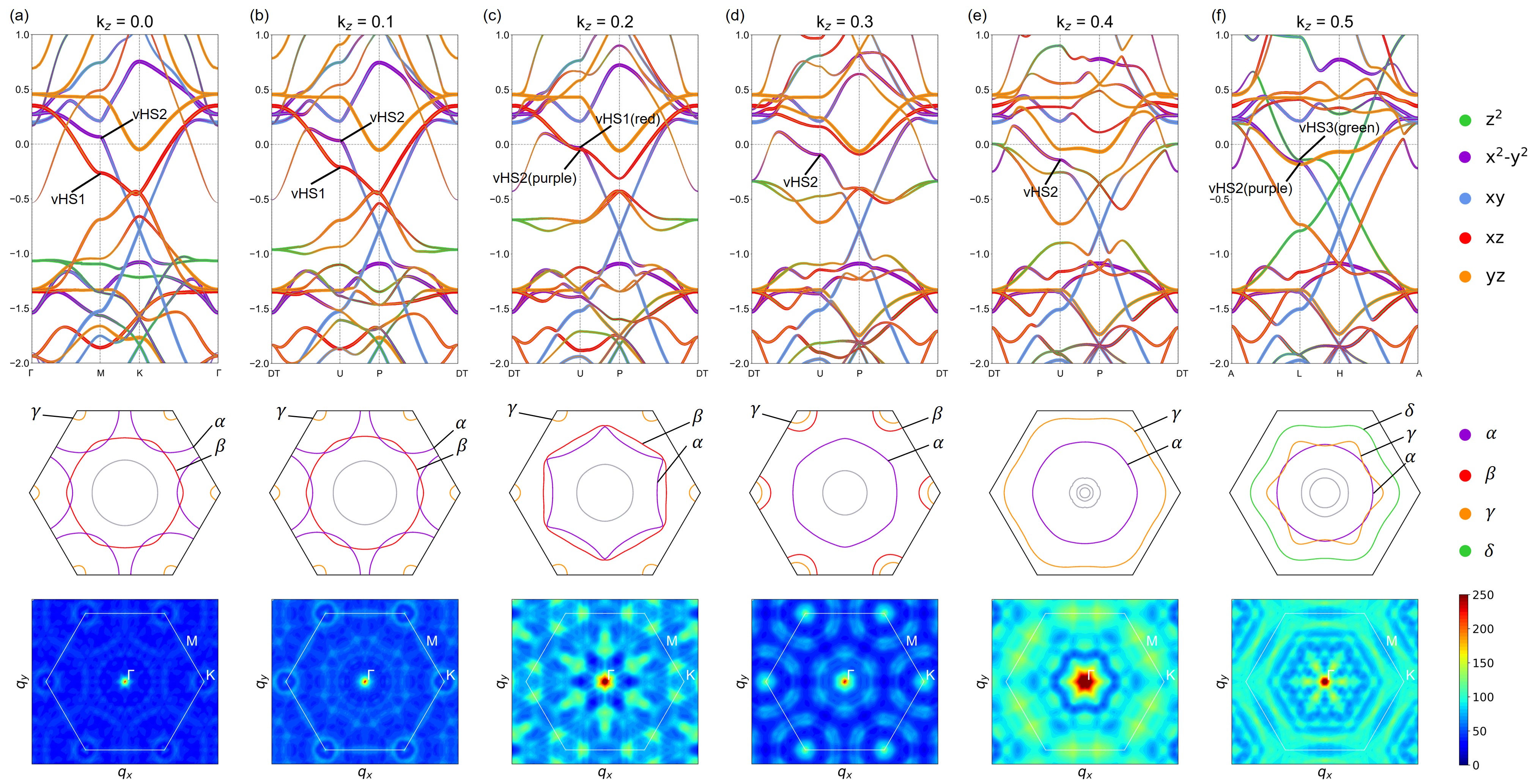}
\caption{(a)-(f) The orbital-projected electronic band structures (top
panel) and 2D Fermi surface (middle panel), and the nesting function (bottom
panel) of FeGe along the $DT(0,0,k_{z})-U(\frac{1}{2},0,k_{z})-P(\frac{1}{3},%
\frac{1}{3},k_{z})-DT(0,0,k_{z})$ path in the $k_{z}$=0, 0.1, 0.2, 0.3, 0.4,
and 0.5 planes. For the orbital-projected band structures, the circle size
shows the relative portion of each orbital. For the 2D Fermi surface, the
pockets $\protect\alpha $, $\protect\beta $, $\varGamma$, and $\protect%
\delta $ represent the pockets dominated by a combination of Fe-$d_{xy}$ and
$d_{x^{2}-y^{2}}$ , Fe-$d_{xz}$, Fe-$d_{yz}$, and Fe-$d_{z^{2}}$ orbital
characters, respectively. }
\label{kz-bandandfm}
\end{figure*}

As shown in Fig. \ref{kz-bandandfm}(a), for $k_{z}=0$ plane, the Fermi
surface pocket formed by the $d_{xz}$-dominant energy band (labeled as $%
\beta $) presents a hexagon parallel to the BZ (hereafter called the
h-hexagon), and the maximum of nesting function is located at $K$ point
(bottom panel of Fig. \ref{kz-bandandfm}(a)). Differently, the Fermi surface
structure in AV$_{3}$Sb$_{5}$ \cite{135_EPC1,135_nest} is v-hexagon, i.e. a
hexagon with a difference of 60$^{\circ }$ rotation from the BZ edge
direction, where the peak of nesting function is located at $M$ point. It is
worth mentioning that in Fig. \ref{susall}, besides the maximum of total
nesting function located at the $K$ point, there are also peaks along the $%
\varGamma-M$ direction. Therefore we carefully analyse the $k_{z}$
momentum-dependent evolution of band structure, Fermi surface and the
nesting function, and show the results in Fig. \ref{kz-bandandfm}(a-f). As $%
k_{z}$ increases from 0 to 0.2, vHS1 gradually approaches the Fermi surface,
and the Fermi surface pocket $\beta $ gradually changes from h-hexagon to
v-hexagon as shown in Fig. \ref{kz-bandandfm}(a-c). The $d_{x^{2}-y^{2}}$/$%
d_{xy}$\ bands forming the pocket $\alpha $ and containing vHS2 shift down
as $k_{z}$ increases, and the shape of the pocket $\alpha $ changes from a
v-hexagonal shape (see Fig. \ref{kz-bandandfm}(c)) to a circle (see Fig. \ref%
{kz-bandandfm}(d)), and then gradually to an h-hexagonal shape (see Figs. %
\ref{kz-bandandfm}(e) and (f)) as $k_{z}$ increases. For $k_{z}$=0.5 plane,
the Fermi surface evolved to contain three different h-hexagonal-shaped
pockets. 
As can be seen from Fig. \ref{kz-bandandfm}, although the shape of the Fermi
surface changes with $k_{z}$, pockets $\alpha $ and $\beta $ both retain
their hexagonal-like shape for a wide range of $k_{z}$ values, suggesting a
large contribution to the Fermi surface nesting.

For the $k_{z}$= 0 and 0.1 and planes, the maximum of nesting function are
located at or near the $K$ point, as shown in Fig. \ref{kz-bandandfm}(a) and
(b). While in the $k_{z}$=0.2 plane, the nesting function shows an
enhancement along the $\varGamma-M$ direction in the momentum space, as
shown in Fig. \ref{kz-bandandfm}(c), which is originate from the nesting of
v-hexagonal pockets $\alpha $ and $\beta $ in Fig. \ref{kz-bandandfm}(c). As
$k_{z}$ continues to shift to around 0.4, the shape of the pockets $\alpha $
and $\beta $ changes again from a v-hexagon to an h-hexagon, with the
maximum of nesting function gradually moving away from the $\varGamma-M$
direction to the $\varGamma-K$ direction. For $k_{z}$=0.5 plane, the
h-hexagonal pockets around $\varGamma$ with similar area to the BZ plane
lead to the maxima of the nesting function near the $\varGamma$ point along
the $\varGamma-K$ direction, as shown in Fig. \ref{kz-bandandfm}(f).

The difference between the Fermi surface and nesting functions of AV$_{3}$Sb$%
_{5}$ and that of FeGe origin from their different crystal structures. In AV$%
_{3}$Sb$_{5}$ (A = K, Cs, Rb), due to the presence of the $A$\ layers, the
distance between the interlayer V atoms is at least 9.308 $\mathring{A}$. In
FeGe, on the other hand, the minimum distance between nearest-neighboring
interlayer Fe atoms is 4.041 $\mathring{A}$, since in FeGe only Fe-Ge$^{in}$
and Ge$^{out}$ layers are alternately arranged. Thus, the hopping parameters
of Fe-3$d$ orbitals in the z-direction are significantly larger than those
of V-$d$ orbitals, and the bands near the Fermi surface in FeGe, which are
mainly contributed by Fe-3$d$ orbitals, have strong 3D features.

In tight-binding models of kagome lattice \cite%
{kagome-03,kagome-06,phase2,kagome3band} and AV$_{3}$Sb$_{5}$ \cite%
{135_PRM1,135_EPC1,135_nest}, the vHSs near the Fermi level could induce the
peak of nesting at $M$ point. Meanwhile, in AV$_{3}$Sb$_{5}$, the phonon
instability \cite{135_PRM1,135_EPC1} and the CDW transition \cite%
{135-05,135-09,135-11,135_nature2,135_PRL2,135_nature3}\ at $M$ point\
driven by Fermi surface nesting\ is suggested. However, in FeGe, where the
Fermi surface has strong 3D feature, the vHSs are not always near the Fermi
energy level. For example, vHS1 is close to the Fermi level ($\pm $0.1 eV)
only in the small range of $k_{z}$=0.183-0.267, which is 16.8\% of the BZ.
Meanwhile, the fraction of $k_{z}$ when vHS2 and vHS3 are close to the Fermi
level are 53.4\% and 13.4\%, respectively. Our numerical results show that
the ratio of nesting function contributed by vHS1, vHS2, and vHS3 around the
Fermi level to the total nesting function are 11.02$\%$, 20.44$\%$ and 7.53$%
\%$, respectively. It means that unlike AV$_{3}$Sb$_{5}$ \cite%
{135_PRM1,135_EPC1,135_nest}, vHSs near the Fermi surface is not the most
important factors of the nesting function in FeGe.

\begin{figure}[tb]
\centering\includegraphics[width=0.48\textwidth]{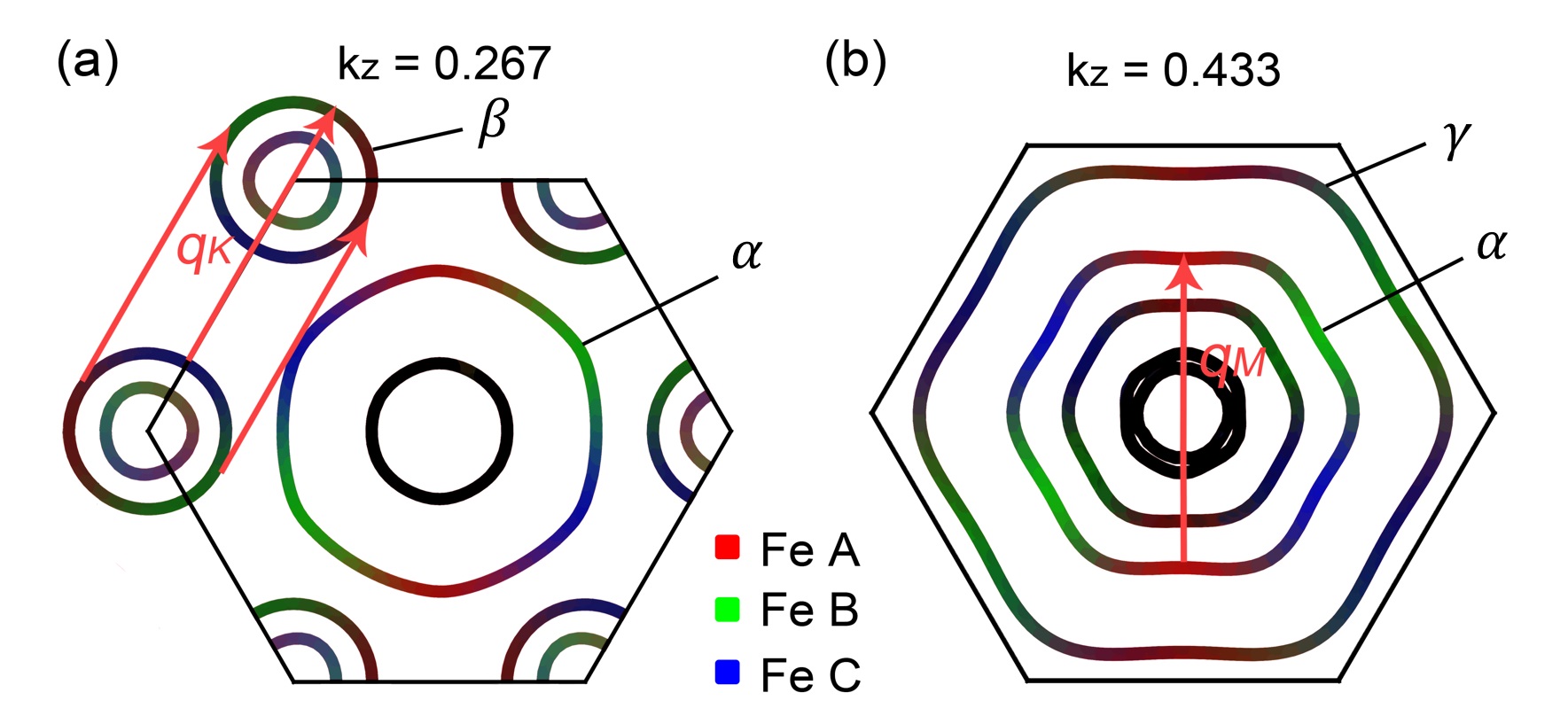}
\caption{The 2D Fermi surface in $k_{x}-k_{y}$ plane with (a)$k_{z}$=0.267
and (b)$k_{z}$=0.433. The sublattice characters of the Fermi level states
including three different Fe$_{A}$, Fe$_{B}$, and Fe$_{C}$ sites, are marked
in red, green and blue, respectively. The red arrows indicate the nesting
vectors $q_{K}$=(1/3,1/3,0) and $q_{M}$=(1/2,0,0).}
\label{metrix-element}
\end{figure}

In FeGe, the maximum of nesting function is at the $K$ point, which does not
correspond to the observed CDW wave vector at $M$ point, indicating the
interaction plays an important role in CDW transition \cite%
{mahan,green_function}. Meanwhile, the theoretically calculated phonon
spectrum remain positive \cite{FeGe_exp1,FeGe_cdw2,FeGe_EPC}. The
electron-phonon interaction is generally weakly momentum-dependent except
for the singular behaviour in non-traditional superconductivity \cite{NC2014}%
. Therefore, we believe that the conflict between the nesting function at K
point and the CDW transition at M point does not originate from
electron-phonon interaction. Therefore, we analyze the local
electron-electron correlation interaction, which is of substantial
importance in 3$d$ electron systems \cite{3dU}.

Based on DFT calculations, we obtain the wave functions $\psi _{n,\mathbf{k}+%
\mathbf{q}}\left( \mathbf{r}\right) $ and $\psi _{n,\mathbf{k}}\left(
\mathbf{r}\right) $, which are electronic Bloch states at the Fermi level
connected by the vector $\mathbf{q}$. We expand the distribution of $\psi
_{n,\mathbf{k}+\mathbf{q}}\left( \mathbf{r}\right) $ and $\psi _{n,\mathbf{k}%
}\left( \mathbf{r}\right) $ to the basis set of atomic orbitals in real
space. We find that when $\mathbf{q}=$ $\mathbf{q}_{K}(1/3,1/3,0)$, the wave
functions $\psi _{n,\mathbf{k}+\mathbf{q}}\left( \mathbf{r}\right) $ and $%
\psi _{n,\mathbf{k}}\left( \mathbf{r}\right) $ are with unequal predominant
sublattice occupancy. Meanwhile, when $\mathbf{q}=$ $\mathbf{q}_{M}(1/2,0,0)$%
, the wave functions $\psi _{n,\mathbf{k}+\mathbf{q}}\left( \mathbf{r}%
\right) $ and $\psi _{n,\mathbf{k}}\left( \mathbf{r}\right) $ are mainly
distributed from the same Fe site. We take a representative $k_{z}$=0.267
plane as an example to show the above mentioned results for the nesting
vector $\mathbf{q}_{K}$ in Fig. \ref{metrix-element}(a), with the characters
of three sublattice Fe$_{A}$, Fe$_{B}$, and Fe$_{C}$ indicated in red, green
and blue, respectively. The Fermi surface contours of $\beta $ pockets
coincide when shifted along the nesting vector $\mathbf{q}_{K}(1/3,1/3,0)$,
as shown in red arrows of Fig. \ref{metrix-element}(a), resulting in the
peak of nesting function at $K$ point. However, the nesting vector $\mathbf{q%
}_{K}$ connects Fermi surface points with mainly different sublattice
occupation, as shown in Fig. \ref{metrix-element}(a). It is worth mentioning
that, due to the locality of Coulomb correlation, the electron-electron
correlation interaction is diagonal in the index of atomic sites. It means
that the susceptibility is suppressed regardless of the peak of nesting
function at $K$ point \cite{green_function}. Meanwhile, we demonstrate the
results of the nesting vector $\mathbf{q}_{M}$ by $k_{z}$=0.433 plane in
Fig. \ref{metrix-element}(b). There are nested Fermi surfaces along $\mathbf{%
q}_{M}(1/2,0,0)$ connecting the opposite edges of Fermi pocket $\alpha $, as
shown in red arrow of Fig. \ref{metrix-element}(b). It can be seen that the
wave functions $\psi _{n,\mathbf{k}+\mathbf{q}}\left( \mathbf{r}\right) $
and $\psi _{n,\mathbf{k}}\left( \mathbf{r}\right) $ connected by the vector $%
\mathbf{q}_{M}$ are dominated by same sublattice occupancy as mentioned
above, leading to enhanced susceptibility at the $M$ point. Therefore,
similar with the sublattice mechanism in superconductors \cite%
{kagome3band,sub-prl}, the CDW instability at $M$ point is considered to be
derived by the local electron correlation. Since the electronic instability
can significantly affect phonons \cite{mahan,green_function}, it may explain
the experimentally observed phonon anomalies \cite{FeGe_exp1,FeGe_EPC}.

In summary, based on DFT calculations, we comprehensively investigated the
Fermi surface nesting and the microscopic origin of the CDW order in the
kagome magnetic metal FeGe. Our results indicate that the energy bands and
Fermi surfaces of FeGe vary significantly with $k_{z}$, and the maximum of
nesting function is at the $K$ point instead of the CDW vector at $M$ point.
We find that the susceptibility at the $K$ point is significantly suppressed
due to the sublattice interference mechanism \cite{kagome3band,sub-prl}, on
the other hand the CDW instability at the $M$ point is enhanced, which
indicates that the electron correlation plays an indispensable part in the
CDW transition.

\section{Acknowledgement}

This work was supported by the NSFC (No. 12188101, 11834006, 12004170,
11790311, 51721001), Natural Science Foundation of Jiangsu Province, China
(Grant No. BK20200326), and the excellent programme in Nanjing University.
Xiangang Wan also acknowledges the support from the Tencent Foundation
through the XPLORER PRIZE.

\newpage

\section{Supplemental Materials}

\begin{figure}[h]
\centering\includegraphics[width=0.48\textwidth]{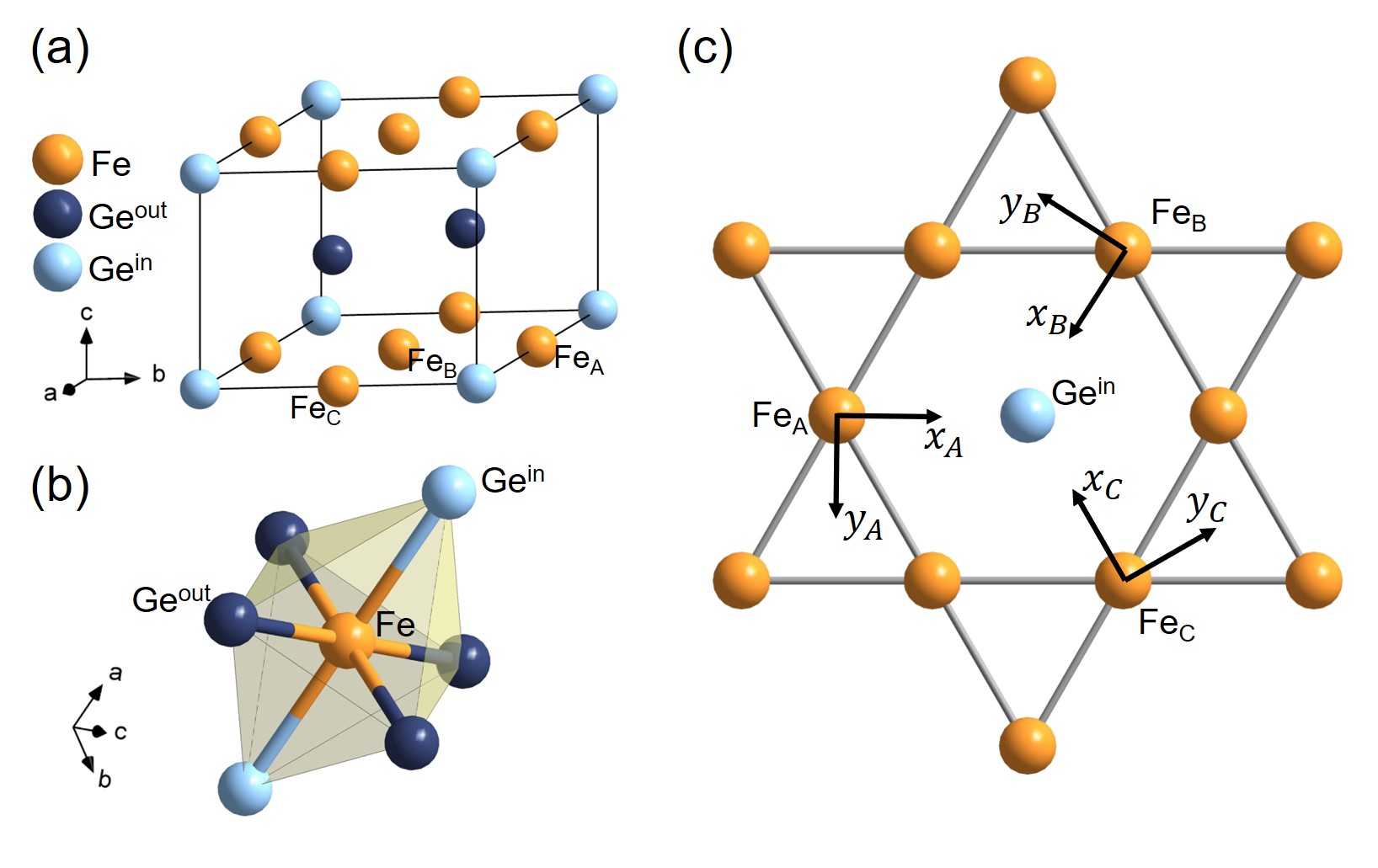}
\caption{(a)The crystal structure of FeGe. The light blue, dark blue, and
orange spheres represent Ge$^{in}$, Ge$^{out}$, and Fe atoms, respectively.
The Fe atoms located at three different sites are labeled as Fe$_{A}$, Fe$%
_{B}$, and Fe$_{C}$. (b) The local structure of the FeGe$_{6}$ octahedron.
The octahedron includes two Ge$^{in}$ atoms and four Ge$^{out}$ atoms. (c)
Our local coordinate for each Fe atom. The local x-axis always points to the
nearest-neighbor Ge$^{in}$ atom, and the z-axis of each local coordinate is
perpendicular to the paper surface.}
\label{structure}
\end{figure}


\subsection{Crystal Structure}

Hexagonal FeGe is an intermetallic compound of the CoSn structure and
crystallize into the $P6/mmm$ (No. 191) space group\cite{FeGe_1963}. As
shown in Fig. \ref{structure}(a), there are two distinct types of Ge atoms
in a unit cell, labeled Ge$^{in}$ (site 1a) and Ge$^{out}$ (site 2d)
respectively, depending on whether they are on the same layer as Fe atoms.
In the Fe-Ge$^{in}$ plane, three Fe atoms at different sites (noted as Fe$%
_{A}$, Fe$_{B}$, and Fe$_{C}$ in Fig. \ref{structure}(a)) form Kagome
lattices, and Ge$^{in}$ atoms are located in the center of the hexagons. Ge$%
^{out}$ atoms compose honeycomb structures above and below the Fe-Ge$^{in}$
plane.

The local structure of the FeGe$_{6}$ octahedron is shown in Fig. \ref%
{structure}(b). It can be seen that each Fe atom is surrounded by six Ge
atomic octahedrons, including two Ge$^{in}$ atoms and four Ge$^{out}$ atoms.
In the $O_{h}$ crystal field, ortho-octahedral structure leads to $%
t_{2g}-e_{g}$ energy splitting. Here the octahedron is distorted and induces
further splitting of the five Fe-3$d$ orbitals.

\begin{figure}[tb!!]
\centering\includegraphics[width=0.48\textwidth]{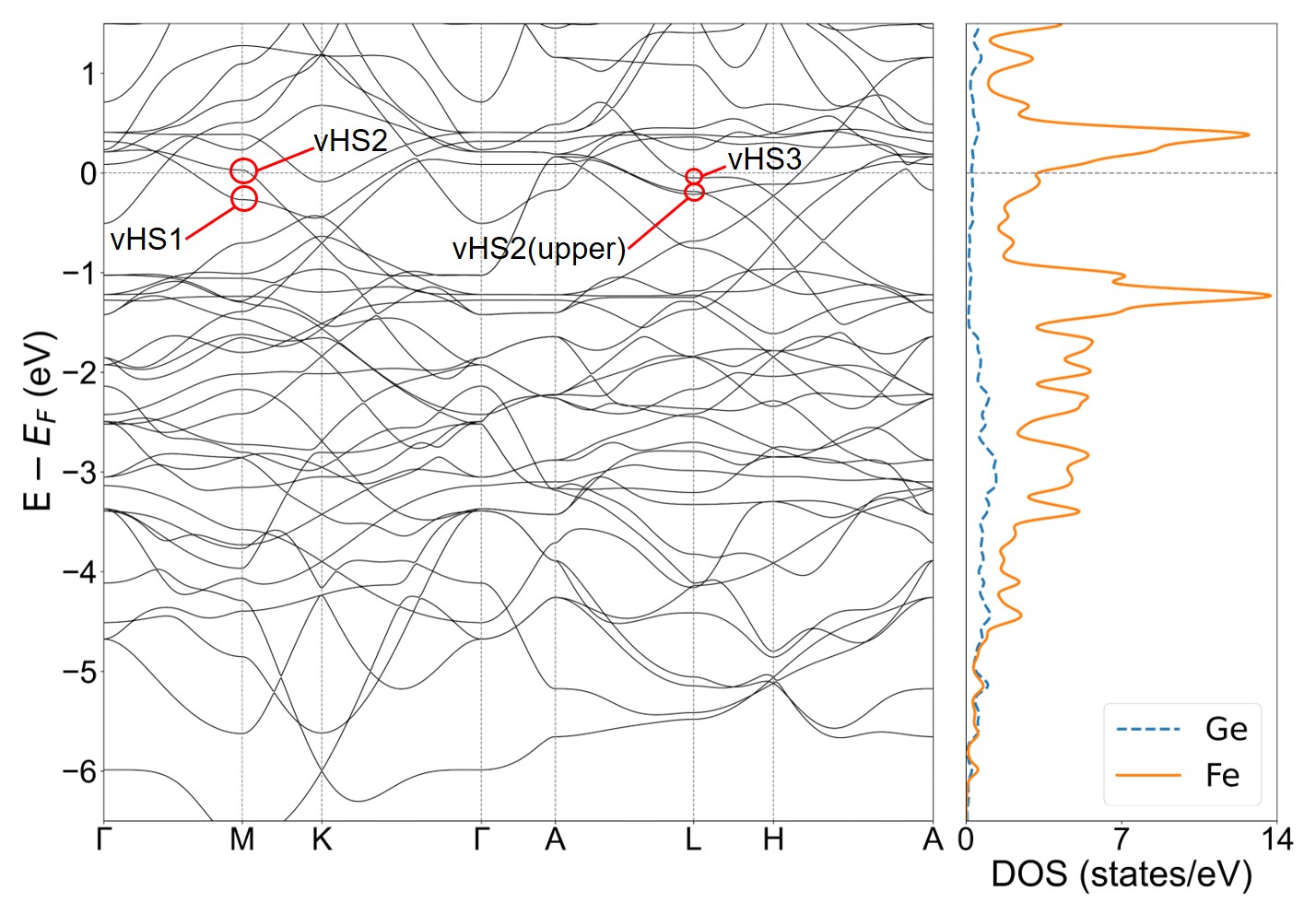}
\caption{Band structure (left panel) and DOS (right panel) of A-type AFM
FeGe. The Fermi level is aligned to 0eV. The DOS of Fe-3$d$ orbitals and Ge-4%
$p$ orbitals are indicated by the solid orange and dashed blue lines,
respectively.}
\label{dos}
\end{figure}

\subsection{Band structure and density of states}

We perform LSDA calculation for FeGe based on the experimental A-type AFM
ground state \cite{FeGe_1988}, and show the band structure and the DOS in
Fig. \ref{dos}. As shown in Fig. \ref{dos}, it is clear that the Fe-3$d$
orbitals dominate the DOS around $E_{F}$ (-2 to 2 eV relative to $E_{F} $),
while the Ge-4$p$ orbitals are mainly located between -6.0 and -2.0 eV. In
the AFM phase, there is an upshift of the spin minority bands due to the
exchange splitting induced by the ordered moment. Two peaks located near 0.4
and -1.3 eV correspond to the flat bands of the spin minority and spin
majority states, respectively. The contribution of the Fe-3$d$ and Ge-4$p$
orbitals is relatively close in the range from -6 to -4 eV, suggesting
hybridization between the Fe-3$d$ and Ge-4$p$ orbitals.

\bibliography{fege-ref-0211}

\end{document}